\documentclass[prb,10pt,twocolumn,showpacs,showkeys,superscriptaddress]{revtex4-1}

\usepackage{graphicx}% Include figure files
\usepackage{dcolumn}% Align table columns on decimal point
\usepackage{bm}% bold math
\usepackage{amssymb}
\usepackage{amsmath}
\usepackage{array}
\usepackage{rotating}
\usepackage{color}
\usepackage{tabularx}
\usepackage{hyperref}
\usepackage[UKenglish]{babel}
\usepackage[utf8x]{inputenc}
\usepackage[T1]{fontenc}

\begin{document}

\title{Optical Magnetic Lens: towards actively tunable terahertz optics}

\author{Georgii Shamuilov}
\affiliation{%
Department of Physics and Astronomy, Uppsala University, L\"{a}gerhyddsv\"{a}gen 1, Uppsala, 75120, Sweden}%

\author{Katerina Domina, Vyacheslav Khardikov}
\affiliation{School of Radio Physics, V. N. Karazin Kharkiv National University, 4, Svobody Square,
Kharkiv 61022, Ukraine}
\affiliation{Institute of Radio Astronomy of National Academy of Sciences of Ukraine, 4, Mystetstv Street,Kharkiv 61002, Ukraine}

\author{Alexey Y. Nikitin}
\email[corresponding author: ]{alexey@dipc.org}
\affiliation{Donostia International Physics Center (DIPC), 20018 Donostia-San Sebasti\'an, Spain}
\affiliation{IKERBASQUE, Basque Foundation for Science, 48013 Bilbao, Spain}
\affiliation{CIC nanoGUNE BRTA, Tolosa Hiribidea 76, E-20018 Donostia-San Sebasti\'an, Spain}

\author{ Vitaliy  Goryashko}
\email[corresponding author: ]{vitaliy.goryashko@physics.uu.se}
\affiliation{%
Department of Physics and Astronomy, Uppsala University, L\"{a}gerhyddsv\"{a}gen 1, Uppsala, 75120, Sweden}%

\date{\today}

\begin{abstract}
As we read this text, our eyes dynamically adjust the focal length to keep the line image in focus on the retina. Similarly, in many optics applications the focal length must be dynamically tunable. In the quest for compactness and tunability, flat lenses based on metasurfaces were introduced. However, their dynamic tunability is still limited because their functionality mostly relies upon fixed geometry. In contrast, we put forward an original concept of a tunable Optical Magnetic Lens (OML) that focuses photon beams using a subwavelength-thin layer of a magneto-optical material in a non-uniform magnetic field. We applied the OML concept to a wide range of materials and found out that the effect of OML is present in a broad frequency range from microwaves to visible light. For terahertz light, OML can allow 50\% relative tunability of the focal length on the picosecond time scale, which is of practical interest for ultrafast shaping of electron beams in microscopy. The OML based on magneto-optical natural bulk and 2D materials may find broad use in technologies such as 3D optical microscopy and acceleration of charged particle beams by THz beams.
\end{abstract}

%\pacs{...}  

\date{\today}
\maketitle

The lens as a tool for focusing transmitted light has been around for four thousand years~\cite{King2003history}. It imprints a proper phase shift onto a light wavefront making the wavefront converging. Conventional optical components (lenses, waveplates, prisms) are \textit{optically thick}~\cite{Saleh2019},
%(transparent materials with a high contrast of indices of refraction are lacking since a higher index of refraction implies lower transparency because of the Kramers-Kronig relations)
and rely on their geometry to imprint required phase shifts by means of the difference in refractive indices. This approach faces a fundamental limitation: the lack of transparent materials with a high contrast of indices of refraction (a higher index of refraction implies lower transparency because of the Kramers-Kronig relations).

% I insist on this wordplay =)
In contrast, a new field of \textit{planar} or \textit{flat optics} has been thriving for the past decade. The concept consists in imprinting abrupt, controlled phase shifts onto transmitted light by a 2D array of subwavelength-thin nanoresonators, \textit{a metasurface}~\cite{Yu2011,Huang2012,Genevet2012,Aieta2012,Chen2012,Pors2013,Yu2014,capasso2018future}. Thus, planar optical components can be made \textit{nanometre thin} and comply with industrial lithography fabrication. 

One of the desired functionalities of both conventional and planar lenses is the active tunability of focal length: think of the eye. Nature's solution realised in mammals' eyes is to tune the focal length by changing the curvature of the lens with the ciliary muscle and by employing a slight gradient of the index of refraction~\cite{coleman1970unified}. A number of eye-inspired approaches and metasurface-based methods have been demonstrated  using mechanical or electric control~\cite{arbabi2018mems,kamali2018review,She2017tunable,she2017large,She2018,Kamali2016,Ee2016,She2018adaptive,Colburn2017,Sautter2015}. Meanwhile, ultrafast and wide active tunability is still challenging~\cite{kamali2018review}. 
\begin{figure}[b]
\centering
\includegraphics[width = \linewidth]{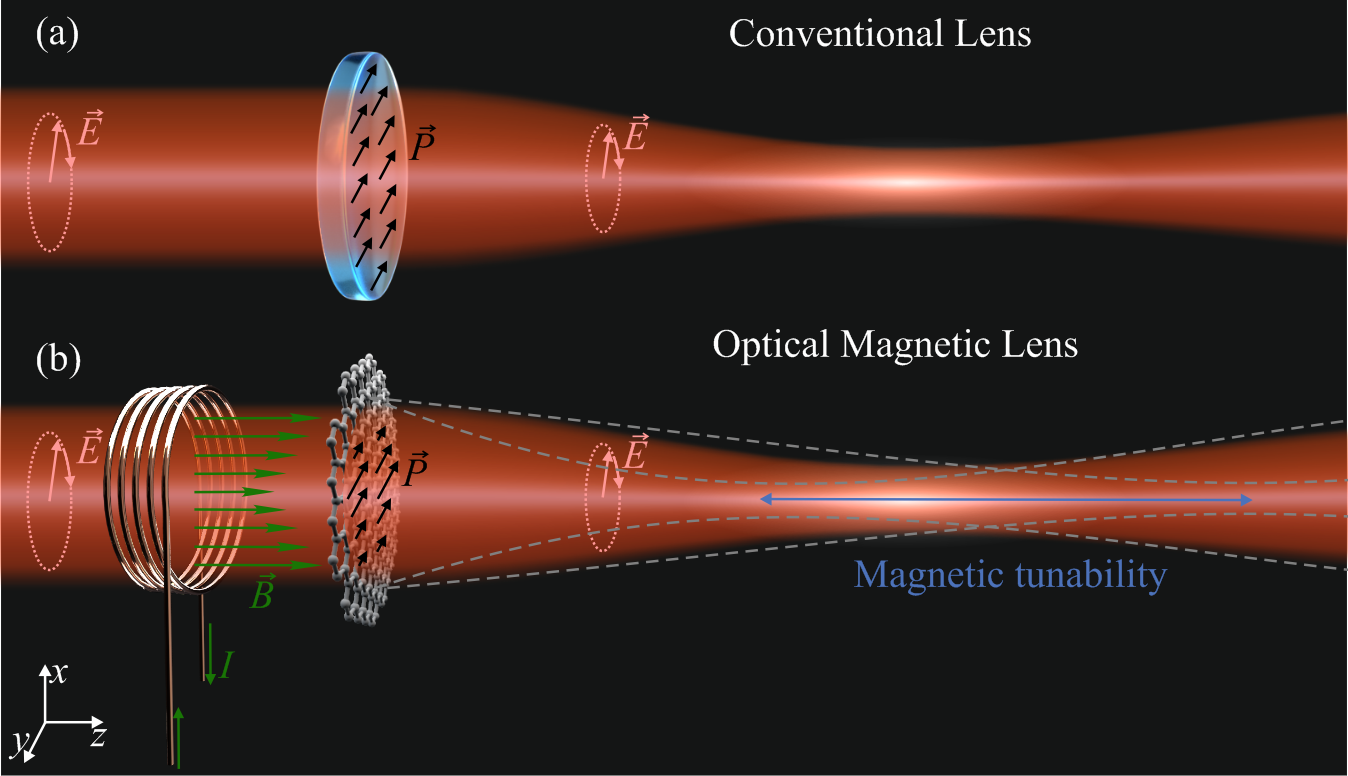}
\caption{(colour online) 
Focusing of a photon Gaussian beam with an electric field $\vec E$ by a lens. Both the geometry and medium polarisation $\vec P$ of  refractive lens \textbf{(a)} govern the cumulative phase shift of the transmitted beam wavefront. In the OML \textbf{(b)}, the phase shift is solely due to non-uniform $\vec P$, which however depends on the transverse profile of the magnetic field $\vec B(\vec r)$ provided by the coil. Thus, in the OML the wavefront can be controlled via $\vec B(\vec r)$. Dashed lines show the envelopes of the focused photon beam for different magnetic field curvatures and the blue arrow indicates the change of the waist location.}
\label{fig:scheme}
\end{figure}

At the same time, actively tunable lenses have been used for around a century in electron microscopy to focus \textit{charged particle beams} by spatially non-uniform  magnetic fields. However, magnetic focusing does not apply to chargeless photon beams. In this Letter, we put forward an original concept of an Optical Magnetic Lens (OML) that focuses \textit{photon beams} using a subwavelength-thin layer of a magneto-optical material immersed into a \textit{non-uniform magnetic field}. We set forth the physics of the OML and exemplify its performance in different frequency bands with bulk and 2D materials. 

The OML features \textit{tunability} of the focal length via changing the strength or curvature of the magnetic field. Specifically, the wavefront of an optical beam incident onto the OML receives a phase shift according to the transverse distribution of the magnetic field strength (Fig.\ref{fig:scheme}). The effect is the most profound in the vicinity of cyclotron resonance in the chosen material, with a phase shift up to one rad, resulting in cm-scale focal distances. %The OML therefore allows for flexible tunability.

\section*{Brief theory}

\begin{figure*}[tb]
\centering
\includegraphics[width=0.95\linewidth]{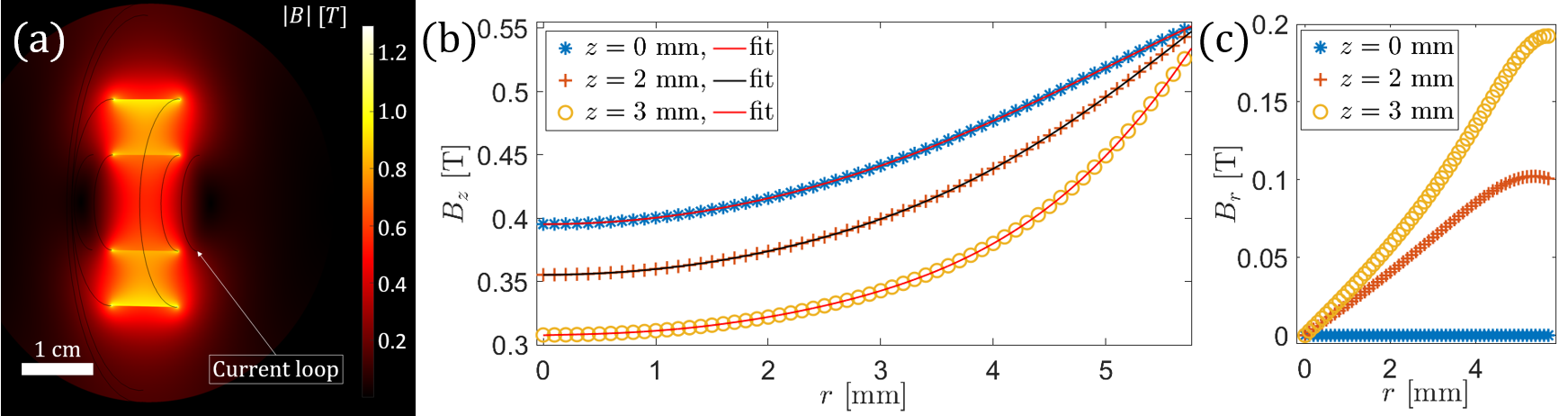}
\caption{(colour online) Simulated spatial distribution of the magnetic field of a commercially available neodymium ring magnet. (a) Magnetic flux density (norm) distribution, a current loop for correction is indicated as an example; (b) The relevant $z$-component of the magnetic field at different positions along $z$, fitted with 6th-order polynomials; corresponding field curvatures $R_c$ are 8.4, 10.3 and 13.6 mm, respectively; (c) Transverse component of the field at the same $z$-positions. }
\label{fig:experiment}
\end{figure*}

To illustrate our concept, let us examine the transformation of a Gaussian optical beam by the OML. Consider a subwavelength-thin, infinitely wide, flat layer of an isotropic medium, supporting free charges. The layer is immersed into an axially symmetric, static, non-uniform magnetic field. The magnetic field is normal to the layer and its strength varies quadratically with distance in the medium [such as the field of a coil or ring, see Figs.~\ref{fig:scheme}, \ref{fig:experiment}, Eq.~\eqref{eq:mag_field_profile} and Supplementary (C)].  Let the origin of a Cartesian coordinate system $(x,y,z)$ be aligned with the extremum of the magnetic field in the layer and the $z$-axis being normal to the layer. A Gaussian optical beam is normally incident onto the layer with the beam waist positioned in the layer: $ \vec E = \vec E_0 \mathrm{e}^{-(x^2+y^2)/w_0^2}$.    
Here, $\vec E_0$ and $w_0$ are the amplitude and waist, respectively.
A sinusoidal behaviour in time at frequency $\omega$ is assumed. 

We treat the layer as an anisotropic medium with its charge carriers oscillating in the combined optical and static magnetic field.  Solving the equations of motion for charge carriers in the layer and relating the electric current to the electric field, we find the  tensor of dielectric permittivity [see~\cite{stix1992waves} and Supplementary (D)]
\begin{equation}\label{eq:tensor_form}
    \hat\varepsilon = 
    \begin{pmatrix}
         1 + \varepsilon_\perp  & -i\beta                & 0 \\
         i\beta                 &  1 + \varepsilon_\perp & 0 \\
         0                      &   0                    & \varepsilon_\parallel 
    \end{pmatrix}
\end{equation}
where $\varepsilon_\perp$ differs from the convention by unity to shorten coming  derivations. 
The elements of this tensor are functions of $r=\sqrt{x^2+y^2}$ 
because the static magnetic field depends on $r$ as $B = B_0 p(r)$, with $B_0$ being the field amplitude and $p(r)$ the radial profile. For field inhomogeneity small on the wavelength scale, $\lambda\, \mathrm{d}(\log p)/\mathrm{d}r < 1$, with $\lambda$ being the wavelength, the functional form of $\hat\varepsilon$ remains unchanged [see Supplementary (D)].

The electric field of the Gaussian beam propagating through the layer in the positive z direction 
is governed by the inhomogeneous paraxial wave equation~\cite{siegman1986lasers}
\begin{equation} \label{eq:paraxial_equation_vector}
    \bigl(\Delta_\perp + 2ik \partial_z \bigr) \vec E = 
         -k^2\bigl(\hat \varepsilon - \hat I \bigr)\vec E,
\end{equation}
where $\hat I$ is the unit tensor, $k \equiv 2\pi/\lambda = \omega/c$ is the wave number, $c$ is the speed of light, 
$\Delta_\perp$ and $\partial_z$ are the transverse Laplacian and  
the partial derivative along $z$, respectively. 
For left-handed (LH), subscript $+$, and right-handed (RH), subscript $-$,
circularly polarised waves $E_{\pm} = E_x \pm i E_y $,
the paraxial wave equation splits and takes on the scalar form
\begin{equation} \label{eq:paraxial_equation_scalar}
    \bigl(\Delta_\perp + 2ik \partial_z \bigr) E_\pm = 
    -k^2\bigl(\varepsilon_\perp \mp \beta \bigr) E_\pm.
\end{equation}
Outside the layer, the longitudinal field component $E_z$ can be found 
from the Coulomb law $\vec\nabla \cdot \vec E = 0$.

The reflected Gaussian beam propagating in the negative z direction 
is described by the same paraxial equation as~\eqref{eq:paraxial_equation_scalar} 
with the only difference that the wavenumber $k$ must be replaced by $-k$. 
As usual in electrodynamics, the boundary conditions consist in the continuity of the electric field $E_\pm$ and its derivative with respect to the propagation coordinate $z$. 

Since the layer is subwavelength thin, diffraction can be safely disregarded and the mathematical problem becomes essentially one-dimensional. The analytical solution for reflected and transmitted waves is described in terms of reflection, $\mathcal{R}$, and transmission, $\mathcal{T}$, Fresnel coefficients, respectively. 
Namely, the electric field of the transmitted wave reads as $E_\pm = \mathcal{T} E_{0,\pm} \mathrm{e}^{-r^2/w_0^2}$. For normal incidence, the Fresnel coefficients take on a simple and well known form~\cite{maier2016world}
\begin{equation}\label{eq:Fresnel_coeffs}
     \mathcal{T}_\pm = \frac{1}{1+\alpha_\pm}, \; \mathcal{R_\pm} = \frac{-\alpha_\pm}{1+\alpha_\pm},\;
    \alpha_\pm = - \frac{i\pi d}{\lambda} \bigl(\varepsilon_\perp \mp \beta \bigr),
\end{equation}
under the assumption that $|\alpha_\pm| \ll 1$. 
General formulas for arbitrary incidence angles can be found, for example, in~\cite{Born2013}. 
The important result is that the Fresnel coefficients locally depends on the transverse coordinate via non-uniform magnetic field. 

For a 2D material with conductivity $\sigma_{\pm}$, the parameter $\alpha_\pm$ is simply the normalised conductivity: $\alpha_\pm = (2\pi/c)\sigma_{\pm}$. Furthermore, for a 2D material the coefficients $\mathcal{T}_\pm$ and $\mathcal{R}_\pm$ in \eqref{eq:Fresnel_coeffs} are exact for any value of $\alpha_\pm$. Note that it is common to describe 2D materials by a conductivity tensor but we choose to use the permittivity tensor to unify the notations for 2D materials and thin layers of bulk materials. 

The Fresnel coefficients in Eq.~\eqref{eq:Fresnel_coeffs} depend on a local static magnetic field, $\mathcal{R}_\pm[B(r)]$, $\mathcal{T}_\pm[B(r)]$, thus setting the spatial phase profile of the reflected and transmitted electromagnetic fields of the beam. The {\it inhomogeneous phase shift} $\varphi_\pm = \mathrm{arg}\bigl\{\mathcal{T}_\pm[B(r)]\bigr\}$ in Eq.~\eqref{eq:Fresnel_coeffs} impacts the shape of the transmitted wavefront. In particular, a quadratic profile of the magnetic field  
\begin{equation}\label{eq:mag_field_profile}
p(r) = 1 + \frac{r^2}{R_c^2}, \quad \frac{r}{R_c} \ll 1    
\end{equation}
with $R_c$ being the radius of curvature, gives the focusing effect. To see the focusing explicitly, we compare the phase shift of the OML to that of a conventional lens, %$\delta\varphi$. The latter is given by 
$\delta\varphi=kr^2/(2f)$, where $f$ is the focal length.  To simplify $\varphi_\pm$, we Taylor expand it with respect to $(r/R_c)$  as
\begin{equation}\label{eq:taylor_expansion_phase}
\varphi_\pm = \varphi_\pm|_{r=0} + \frac{1}{2}~ \varphi_\pm''|_{r=0}~\frac{r^2}{R_c^2},
\end{equation}
with $\varphi_\pm''$ being the second derivative of $\varphi_\pm$. Similarly to $\delta\varphi$, the inhomogeneous phase shift of the OML scales quadratically with $r$ [second term in Eq.~\eqref{eq:taylor_expansion_phase}], thus clearly indicating a focusing effect. By comparing the expanded $\varphi_\pm$ with $\delta\varphi$, we find the focal length of the {\it tunable flat OML} for LH and RH circularly polarised waves 
\begin{equation}\label{eq:focal_length}
    f_\pm = \frac{k R_c^2}{\varphi_\pm''}.
\end{equation}

Image formation by the lens is well known in optics and discussed in Supplementary (B) for completeness.

Examine the contents of results~\eqref{eq:Fresnel_coeffs} and \eqref{eq:focal_length}. First, the homogeneous part of the phase shift $\varphi_\pm$ leads to the Faraday rotation of \textit{linearly} polarised light.
Second, within the layer, left and right circularly polarised components of the electric field experience different effective permittivities $\alpha_\pm$, and the focal length~\eqref{eq:focal_length} contains different signs. Thus, the polarisation components have different focal lengths. Third, a comparison with full-wave simulations showed that the solution~\eqref{eq:focal_length} is accurate under a constraint of $w_0 < 2R_c$, more relaxed than the one in Eq.~\eqref{eq:mag_field_profile}. Fourth, the reflectivity of the layer can be high and thus allows for OML operation in reflecting telescope or mirror geometry.

To obtain an explicit expression for $f_\pm$ as a function of parameters of the film, let us proceed to the elements of the tensor $\hat\varepsilon$, Eq~\eqref{eq:tensor_form}.
The charges oscillate around the applied magnetic field with a frequency $\omega_c(r) = \omega_0 \mathcal{M} p(r)$. Here, $\omega_0=qB_0/m_e c$ is the reference (on-axis) cyclotron frequency with 
$m_e$ being the electron mass, $\mathcal{M}=m_e/m^*$ is the mass reduction ratio, $q$ and $m^*$ are the charge and effective mass of the particle, respectively.
In fact, the elements of the tensor $\hat\varepsilon$ correspond to a magnetised plasma (Drude model)~\cite{Nikolskiy1989,bergman2000magnetized,Tymchenko2013} and read
\begin{equation}\label{eq:eps&beta_2DEG}
    \varepsilon_{\perp} = \frac{-\mathcal{A} (\omega - i/\tau)}{\omega \left[ (\omega - i/\tau)^2 -\omega_c^2(r) \right]}, \;
    \beta =\frac{\omega_c(r)~\varepsilon_{\perp} }{(\omega - i/\tau)},\\
\end{equation}
where $\mathcal{A}$ is a material-specific constant [s$^{-2}$] and $\tau$ is the relaxation time. Particular cases with more familiar expressions for $\varepsilon_{\perp}$ and $\beta$ can be found in Supplementary (A). Following Eq.~\eqref{eq:Fresnel_coeffs}, we obtain a simple expression for the focal length for LH and RH circularly polarised waves transmitted through a layer of thickness $d$ as
\begin{equation}\label{eq:result_2DEG}
    f_\pm = \mp \frac{R_c^2}{\mathcal{A} d}~\frac{\omega}{\mathcal{M}\omega_0}~\frac{[(\omega\mp \mathcal{M}\omega_0)^2+(1/\tau)^2]^2}{(\omega \mp \mathcal{M}\omega_0)^2-(1/\tau)^2}.
\end{equation}
This simple result allows one to calculate the focal length of the OML for different materials as illustrated below. Due to the term $(\omega \mp \mathcal{M}\omega_0)^2$ in the denominator in  \eqref{eq:result_2DEG}, $f_\pm$  has a resonant behaviour for one of the polarisations of the optical beam for a given magnetic field orientation. Namely, for $\omega \approx \mathcal{M}\omega_0$ a cyclotron resonance occurs. At the resonance, Eq.~\eqref{eq:result_2DEG} simplifies to $f_\pm = \mp R_c^2/(\mathcal{A}d\tau^2)$.

\section*{Examples of materials}

Consider potential practical realisations of the OML for different frequency bands. %The simplest OML is a thin metal film, in which $\mathcal{M}=1$ and $\mathcal{A}=\omega_p^2$ (squared plasma frequency). Moderate magnetic fields ($B_0\sim 1$~T) limit maximum $\omega_0$ to some 10 GHz while $1/\tau=5.5$ THz for silver~\cite{Blaber2009}. This large  $1/\tau$ makes the cyclotron resonance weak and results in a focal length on the scale of 10~m ($|T|^2 \approx 15\%$). 

Material suitable for magnetic focusing in the microwave range are \textit{magnetic dielectrics}, or \textit{ferrites}, such as Yttrium Iron Garnet (YIG). Instead of charge carriers, there are unpaired spins precessing in the applied magnetic field. 
The functional form of $\hat\varepsilon$, Eq.~\eqref{eq:tensor_form}, and its components remain unchanged. Hence, the result in Eq.~\eqref{eq:result_2DEG} can be applied directly to ferrites, where $\omega_0$ should be understood as the Larmor frequency [see Supplementary (A)]~\cite{pozar2011microwave}. Practical results for the OML in the microwave region are presented in Table~I. A focal length of tens of centimeters  is feasible. 
Ferrite-coated mirrors can potentially be used for \textit{tunable} focusing of quasi-optical microwave beams in fusion experiments, e.g. for plasma probing or electron-cyclotron-resonance heaters~\cite{Alberti2007,ITER}.

\begin{table*}[tb]\label{tab:performance}
\caption{Examples of materials for Optical Magnetic Lens in transmission mode (T) at normal incidence $\phi_i=0$ and in reflection mode (R) at $\phi_i=20^\circ$. For InSb, THz phonon resonances must be avoided.}
\begin{tabularx}{0.97\linewidth}{ X|c|c|c|c|c } %{ X|C|C }
 \hline
 Parameter &				Graphene (T) &	Graphene (R) &	InSb (T) &	InSb (R) &	YIG (T) \\
 \hline
 Light frequency $\omega/2\pi$ &	1~THz  &	1~THz &		3~THz &		2~THz &		50~GHz\\
 Relaxation time $\tau$ & 		0.5 ps &	1 ps &		3 ps &		3 ps &		$\approx0.1~\mu$s \\
 Efficiency ($|T|^2$ or $|R|^2$) &	32\% &		55\% &		27\% &		66\% &		72\% \\
 On-axis field $B_0$ &			0.2 T  &	0.2 T &		2.1 T &		1.7 T &		1.8 T\\
 Field curvature $R_c$ &		0.32 cm  &	0.3 cm &	1.5 cm &	1.5 cm &	70 cm\\
 Film thickness $d$ &			monolayer &	monolayer &	$0.6~\mu$m &	$3~\mu$m &	$0.1~\mu$m \\
 Focal length $f$ &			    8 cm &		40 cm &		16 cm &		40 cm &		51 cm \\  
 \hline
\end{tabularx}
\end{table*}

To operate above microwaves, we need a material with a high mass reduction factor $\mathcal{M}$, $\mathcal{M} \gg 1$. \textit{Doped graphene} is an outstanding candidate for a higher-frequency OML.
We use the semiclassical model to describe doped graphene in magnetic fields~\cite{Ferreira2011}. This model accounts only for intraband transitions, but is valid in a broad range covering the terahertz and mid-infrared bands under the condition $\hbar\omega<2|\mu_c|$. Here, $\mu_c$ is the chemical potential and $\hbar$ is the reduced Planck constant. 

To use Eq.~\eqref{eq:result_2DEG} directly for a graphene sheet with a conductivity $\hat\sigma$, we  approximate graphene by a layer with a finite thickness $d$ and introduce an effective dielectric permittivity tensor $\hat\varepsilon_\mathrm{eff}= (4\pi i/\omega d)~\hat\sigma$~\cite{maier2016world}.
Then, the elements of $\hat\varepsilon_\mathrm{eff}$ assume the form given by Eq.~\eqref{eq:eps&beta_2DEG}. 
As it should be for a 2D material, the dependence on $d$ in $f_\pm$ cancels out. 
For doped graphene, $\mathcal{M}$ is $m_eV_F^2/|\mu_c|$. We see that graphene posses an intriguing possibility of increasing the mass reduction factor $M$ by increasing the Fermi velocity $v_F$ and operating with a small chemical potential $\mu_c$. From the practical points of view this implies that the cyclotron resonance can be reached for lower magnetic fields for the same THz frequency.  Recent experiments in the THz and IR regions show that the Fermi velocity can be engineered by placing graphene on a suitable dielectric substrate~\cite{Hwang2012,Whelan2020}.   Assuming a chemical potential $\mu_c=0.19$~eV and Fermi velocity $V_F=2.5\cdot10^6$ m/s, we estimate $\mathcal{M}\approx 187$ and $\mathcal{A}\approx 1.2\cdot10^{18}$~s$^{-2}$. The remaining parameters are listed in the Table~I. 

\begin{figure}[bt]
\centering
\includegraphics[width=\linewidth]{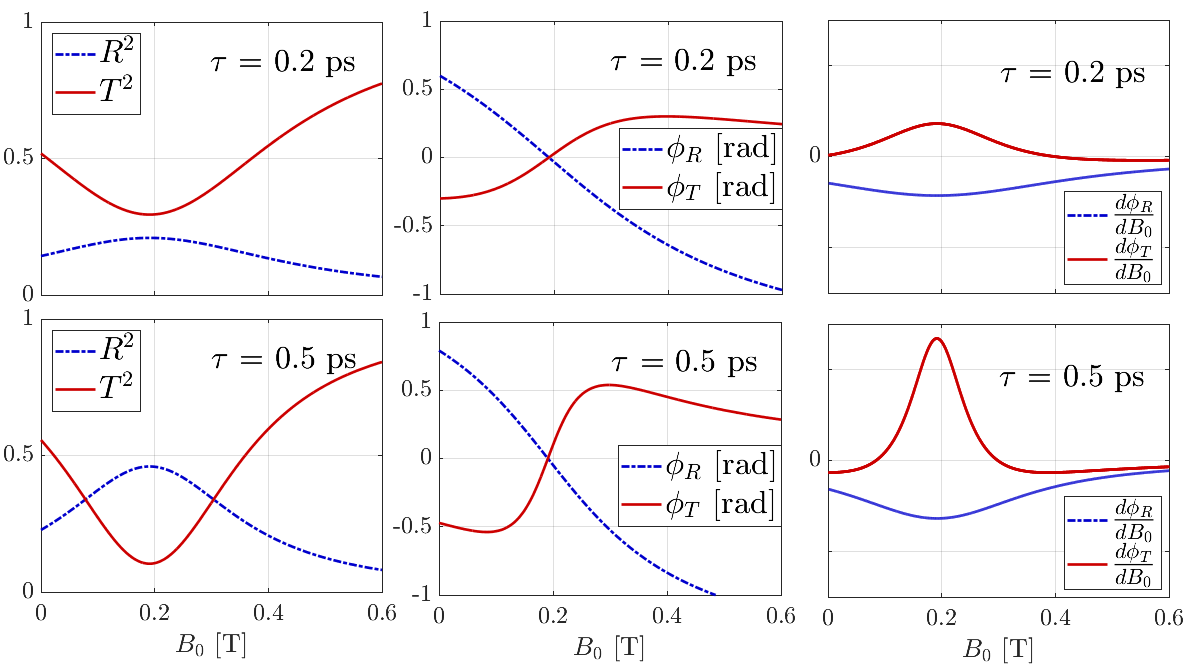}
\caption{(colour online) Transmission, reflection, phase shift and derivative $d\phi/dB_0$ calculated via Fresnel coefficients~\eqref{eq:Fresnel_coeffs} for a graphene layer w.r.t. applied magnetic field $B_0$ (uniform). The chemical potential is 0.19~eV.}
\label{fig:work_point}
\end{figure}

Figure~\ref{fig:work_point} demonstrates a clear resonant behaviour of the transmitivity, reflectivity and phase shifts of a single-layer-graphene OML in a \textit{uniform} magnetic field, typical for the cyclotron resonance. The maximum of the derivative of the phase shift with respect to the magnetic field, $d\phi/dB_0$, suggests an operating point of the OML in a \textit{non-uniform} magnetic field. Namely, for $B_0 \approx 0.2$ T, $d\phi/dB_0$ is maximal and $f$ attains its minimum value in a \textit{non-uniform} $B$ given that other parameters are fixed. The inverse relaxation time of graphene, $1/\tau$, plays the role of the resonance bandwidth: larger values of $\tau$ (high-purity graphene) provide a sharper resonance and thus a larger phase shift (on the order of one radian). At the same time, the OML appears tolerant to smaller $\tau$ values so that the graphene OML does not require high-quality graphene flakes for its reasonable performance.

\begin{figure}[bt]
\centering
\includegraphics[width=0.85\linewidth]{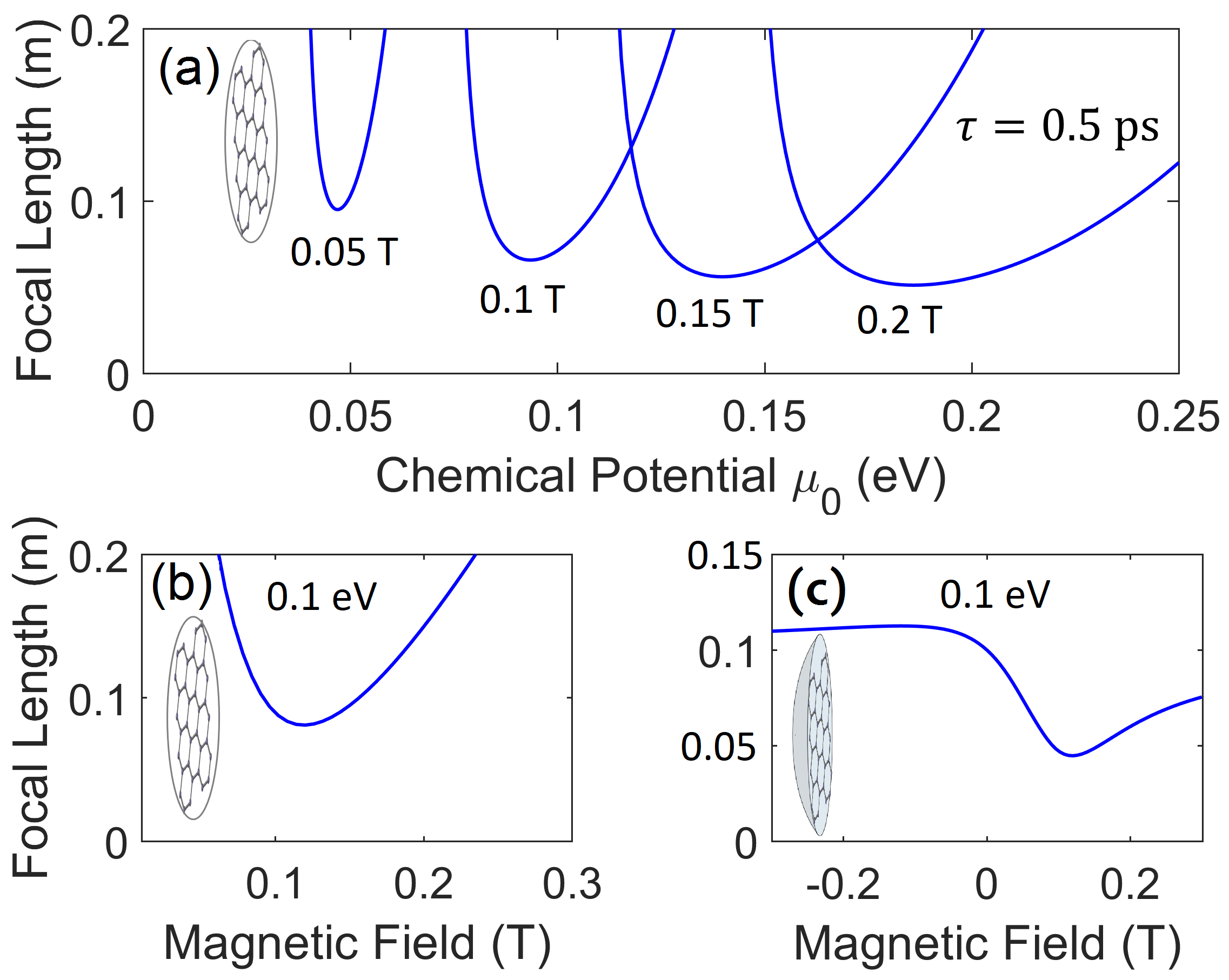}
\caption{(colour online) \textbf{(a)} Focal length of graphene OML vs chemical potential for different values of the magnetic field strength. \textbf{(b)} Focal length of graphene OML vs applied magnetic field. \textbf{(c)} Focal length of a conventional lens \textit{combined} with graphene OML vs applied magnetic field. Here, $\tau = 0.5$~ps, $R_c = 3.2$~mm, $\lambda = 300 \; \mu$m.}
\label{fig:comparison}
\end{figure}

For a LH circular polarisation of the optical beam incident onto graphene OML, we calculate a focal length of some cm with a wide adjustment range given by the field amplitude $B_0$, Fig.~\ref{fig:comparison}b, and curvature $R_c$. Additional active adjustment of the focal length can be done by varying the chemical potential $\mu_c$, Fig.~\ref{fig:comparison}a. Thus, the OML can bring vast tunability into existing THz optics.

We note that only one circular polarisation component undergoes resonant focusing [$\omega \approx \mathcal{M}\omega_0$, see Eq.~\eqref{eq:result_2DEG}] by the OML. Hence, such a lens allows for selective focusing by choosing the direction of the external magnetic field. This effect can be potentially used for polarisation-sensitive detection of THz light.

The OML can also be used in combination with conventional lenses, substantially improving the performance of the latter. As an example, Figure~\ref{fig:comparison}c shows 50\% relative tunability of a  conventional lens, having a fixed focal length $f$ of 10 cm, decorated with the graphene OML. The focal distance of the combined lens can be tuned from around 5 to 12~cm.

To visualise the effect of the OML as well as to cross-check our analytical results, we run full-wave simulations for the particular example of graphene OML. We use commercial software COMSOL Multiphysics. Thanks to the azimuthal symmetry of the problem, the model can be built in 2D to reduce required computation power. The incident Gaussian beam (background field) is defined analytically and the graphene layer is represented as a surface current density given by 2D conductivity tensor~\cite{Ferreira2011,Tymchenko2013,maier2016world}. The presence of the non-uniform magnetic field is included analytically into the conductivity tensor~\cite{Tymchenko2013}. The final field distribution is calculated as the field scattered by the graphene layer.

\begin{figure}[bt]
\centering
\includegraphics[width=0.85\linewidth]{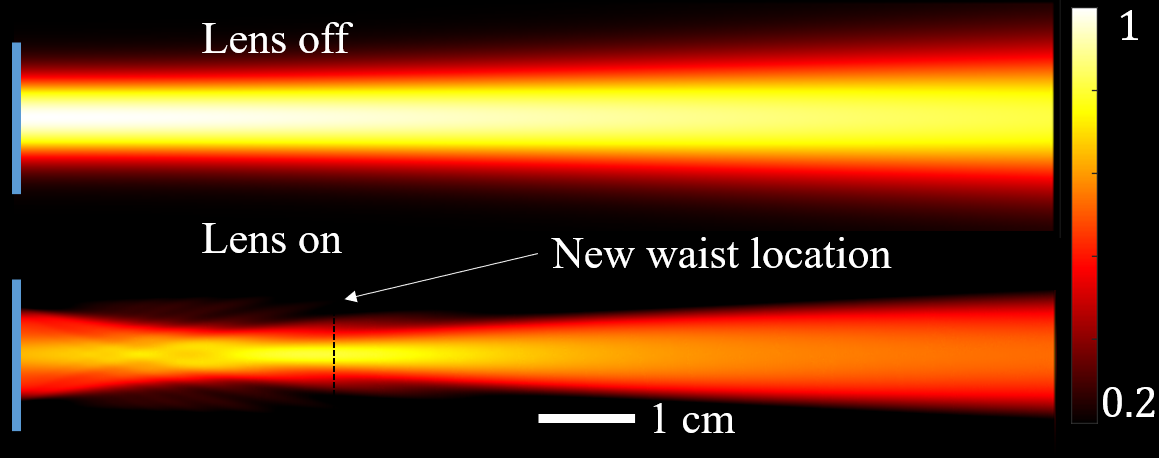}
\caption{(colour online) Top plot: normalised field distribution transmitted through the graphene OML \textit{without} magnetic field (OML ``off''). Bottom plot: normalised field distribution with the OML ``on''.}
\label{fig:comsol_waist}
\end{figure}

The focusing effect is clearly seen in Fig.~\ref{fig:comsol_waist}. If no magnetic field is applied (lens is ``off''), the Gaussian beam diverges in the region to the right from the graphene OML (top plot). In contrast, a new waist of the beam appears (bottom plot), when a profiled magnetic field is applied (lens is ``on''). The focal lengths calculated analytically, 6.5~cm, and numerically, 6.4~cm,  match very well, thus validating our analytical approach, Eqs.~\eqref{eq:tensor_form}-\eqref{eq:result_2DEG}.

In Fig.~\ref{fig:comsol_waist}, for having a sharper image and clearer visual illustration, we partly compensated for lens aberrations by adding a term $-0.8 r^6/R_c^6$ to the magnetic field profile $p(r)$. In the simulation, $\tau = 0.5$~ps, $\mu_0 = 0.1$~eV, $B_0 = 0.09$~T, $R_c = 1.2w_0$, $w_0 = 6\lambda$ with $\lambda = 300 \;\mu$m.

\textit{Semiconductors and their heterostructures} are another important example of materials for the OML. The sophisticated underlying mechanism of charge transport significantly reduces the effective mass of electrons~\cite{tang2006effective,wang2018high}, which can increase OML operating frequencies. 
The highest value of $\mathcal{M}\approx50$ in this class of materials is achieved for indium antimonide (InSb)~\cite{agranovich2012surface}. %Thus, the resonant frequency can be as high as 1.5 THz per tesla of the applied magnetic field $B_0$. 
Assuming parameters tabulated in Ref.~\cite{palik1991handbook}, we calculate $\mathcal{A}=9.26\cdot10^{25}$~s$^{-2}$ and the focal length of about 16~cm (see Table~I). Compared to graphene, tunability in InSb is limited to magnetic field only. Also, InSb exhibits phonon modes in the same frequency range suppressing resonant focusing.

We also anticipate a possibility to use an array of \textit{ferromagnetic nanoresonators} (e.g. TbCo~\cite{Ciuciulkaite2020,RowanRobinson2020}) for the OML at optical frequencies. Operating conditions are similar to those for ferrites discussed above. The focusing effect is weaker than in the THz range (the inhomogeneous phase shift is on the order of 10~mrad for $B_0 \sim 1$~T), but allows for fine-tuning of the focal length if the array is deposited on the surface of a plano-convex lens, similar to the example with graphene in Fig.~\ref{fig:comparison}c.

From a different perspective, the OML effect may impact propagation of electromagnetic waves in space similarly to a gravitational lens. Namely, a wavefront transformation may occur in \textit{cosmic plasma} exposed to non-uniform magnetic fields generated by different massive astrophysical objects, thus affecting divergence of light from a remote source [see Supplementary (A)].

\section*{Experimental considerations}

Let us describe a possible experimental setup for focusing THz light with a graphene-based lens. Reduced to its essentials, the setup can consist of just four key components: (1) the lens itself: a transparent substrate decorated with large-scale graphene fabricated with chemical vapour deposition~\cite{Vlassiouk2013}; (2) a THz source based on optical rectification from, e.g., a zinc telluride crystal~\cite{Salen2019} to generate an optical beam with a spectrum peaking at 1~THz;  (3) a simple ring magnet to set the focal length and shape of the optical beam; (4) a THz beam imager based, for instance, on electro-optical sampling~\cite{Salen2019} or an array of microbolometers~\cite{Liu2020}. In practice, it is advantageous to place an additional thin current loop next to the ring magnet for fine tuning of the magnetic field curvature. 
It turns out that the optical quality of the proposed OML suffers from spherical aberrations if a simple quadratic profile of the magnetic field is applied. The phase shift of the transmitted light $\varphi_\pm$ given by the Fresnel coefficient~\eqref{eq:Fresnel_coeffs} is a complex function of $B$ and hence a complex function of $r$. To correct for the aberrations, the transverse profile of the magnetic field must have not only a quadratic component ($r^2$), but also a component depending on $r^6$. For instance, in the simulation in Fig.~\ref{fig:comsol_waist} the optimal transverse profile of $B$ is $(1 + r^2/R_c^2 - 0.8r^6/R_c^6)$. This profile can be realised in practice by properly choosing the longitudinal position of the graphene layer with respect to the ring magnet plane. In addition, for fine tuning of the magnetic field profile a current loop can be used. 

For typical OML operation the radius of curvature of the magnetic field must be larger than the THz beam waist, $R_c \approx (1.2-1.3) w$. At the same time, for a ring magnet  $R_c$ is usually smaller than the physical radius of the ring $R$, see Fig.~\ref{fig:experiment}. Hence, nearly 100\% transmission of the THz beam through the aperture of the ring is possible since $R \approx 1.5 w$. The typical numerical aperture is 0.1.

Thus, we have four different knobs in the OML magnet design to compensate for spatial aberrations: (i) graphene layer position w.r.t. the ring magnet, (ii) separation between the ring magnet and the current loop, (iii) current loop radius and (iv) the number of windings.

\section*{Discussion}

At THz frequencies, the response time of the graphene-based OML can potentially be as short as a few picoseconds. Though the physical mechanism of the phase shift induced in the OML is resonant and relies on cyclotron resonance, the relaxation time is typically less than a picosecond. That allows for ultrafast tuning.  In practice, the response time will be limited by technical auxiliaries such as the response time of an electromagnetic coil used to create the required magnetic field profile. However, there is a promising solution for tuning the graphene-based OML on the picosecond time scale: to use quasi-half-cycle THz pulses~\cite{Hebling2002,Salen2019} to additionally control the chemical potential, see Fig.~\ref{fig:comparison}a, and correspondingly adjust the focal length. 

In contrast to sinusoidal electromagnetic pulses, quasi-half-cycle pulses maintain their electric field oriented in the preferential direction. Hence, the effect of such pulses on the graphene layer can be thought of as an instantaneous DC voltage. A permanent magnet can be used to preset a desired focal length of the OML and the electric field of an additional quasi-half-cycle THz pulse will modify the chemical potential on the picosecond time scale thus adjusting the focal length. %See Fig.~\ref{fig:comparison}a for focal length tuning via changing the chemical potential for a fixed $B_0$.

In summary, we introduced a concept of the magnetically tunable flat lens. It takes advantage of the resonant magnetic-field-dependent phase shift and features tunability by means of magnetic field control. We applied our model to a wide range of materials (noble metals, semiconductors, graphene, ferrites and nanoparticle arrays), and found out that, with varying efficiency, the OML can be realised in a broad frequency range from microwaves to visible light. Moreover, using other magnetic field profiles, our OML can be reconfigured to operate as another optical component, e.g. as a beam deflector with a linear field profile or a grating with periodic field profile. We anticipate that the OML, based on available magneto-optical bulk and 2D materials, can find wide use in many optoelectronic technologies in a broad spectral range.

%\section*{References}
%\bibliographystyle{ieeetr}
%\bibliography{references_MO_lens}

\section*{Acknowledgements}
We thank Prof. Oleg Kochukhov and Assoc. Prof. Vassilios Kapaklis  (Uppsala University) for fruitful discussions, and Prof. Paolo Vavassori (CIC NanoGUNE).

V.G. acknowledges the support of Swedish Research Council (Vetenskapsr{\aa}det) (grant No.
2016–04593) and A.Y.N. acknowledges the Spanish Ministry of Science, Innovation and Universities (national project MAT2017-88358-C3-3-R) and Basque Government (grant No. IT1164-19).

\section*{Additional information}
Supplementary Information accompanies this paper.

\newpage
\section{Supplementary Materials}
%\maketitle
%\tableofcontents

\subsection{Permittivity tensors for various media}

For convenience of the reader, we include the expressions for elements of the dielectric permittivity tensor $\hat\varepsilon$ [Eq.~(1) in the article] for different materials considered in the paper.

\subsubsection{Plasmonic material}

A thin film made of silver or gold is represented in the same way as magnetised plasma~\cite{Nikolskiy1989,bergman2000magnetized}
\begin{equation}\label{eq:eps&beta_plasma}
\begin{split}
    1+\varepsilon_{\perp} &= 1 - \frac{\omega_p^2~(\omega - i/\tau)}{\omega \left[ (\omega - i/\tau)^2 - \omega_c^2 \right]},\\
    \beta &= \frac{\omega_c~\omega_p^2}{\omega \left[ (\omega - i/\tau)^2 - \omega_c^2 \right]}.
\end{split}
\end{equation}
Here, $\omega$ is the frequency of light, $\omega_p$ is the material's plasma frequency and $\tau$ is the relaxation time. For silver, $\omega_p=$~2321~THz and $1/\tau=$~5.513~THz~\cite{Blaber2009}. Assuming a square-shaped magnetic field profile, as required for focusing, $\omega_c$ reads
\begin{equation}\label{eq:cyclotron_freq}
\omega_c = \frac{qB_0}{m^*c} \left( 1+\frac{r^2}{R_c^2} \right) = \omega_0 \mathcal{M} \left( 1+\frac{r^2}{R_c^2} \right),
\end{equation}
with $\mathcal{M}=m_e/m^*=1$ for electrons in metal. Therefore, the focal length reads
\begin{equation}\label{eq:result_silver_film}
    f_\pm = \mp \frac{R_c^2}{d} \frac{\omega}{\omega_p^2\omega_0} \frac{[(\omega \mp \omega_0)^2+1/\tau^2]^2}{(\omega \mp \omega_0)^2-1/\tau^2} \approx \frac{\pm R_c^2}{d\tau^2\omega_p^2}\frac{\omega}{\omega_0},
\end{equation}
where the approximate value takes place under realistic magnetic fields [$\omega_0=(2\pi)~28$~GHz per 1 T of applied field], so that both $\omega$ and $\omega_0\ll 1/\tau$; $R_c$ is the curvature radius of the magnetic field and $d$ is the film thickness. 

\subsubsection{Ferrite}

When working with non-magnetic materials, relative permeability $\mu=1$, so that it is omitted and the Eq.~(2) of the article contains only the permittivity tensor $\hat\varepsilon$. 
For ferrites, the tensor form is traditionally assigned to $\hat\mu$, whereas $\varepsilon$ has a scalar value ($\varepsilon=15$ for YIG). It is equivalent to write the paraxial wave equation~(2) of the article as
\begin{equation}
    \bigl(\Delta_\perp + 2ik \partial_z \bigr) \vec E = 
         -k^2\bigl(\varepsilon \hat \mu - \hat I \bigr)\vec E,
\end{equation}{}
where $\hat\mu$ has the same form as $\hat\varepsilon$ in Eq.~(1) in the article. Let us write down its elements ready for Eq.~(2) of the article~\cite{pozar2011microwave}
\begin{equation}\label{eq:eps&beta_ferrite}
\begin{split}
    1+\varepsilon_{\perp} &= \varepsilon \left[ 1 - \frac{\omega_M~(\omega_c+i\alpha\omega)}{\omega^2 - (\omega_c+i\alpha\omega)^2} \right],\\
    \beta &= \frac{\varepsilon~\omega~\omega_M}{\omega^2 - (\omega_c+i\alpha\omega)^2}.
\end{split}
\end{equation}
Here, $\omega_M=qM_s/m_ec$ with $M_s$ being the saturation magnetisation ($\omega_M=2\pi\cdot49.8$~GHz for YIG), $\alpha=2\cdot10^{-4}$ (tangent loss angle, YIG) and $\omega_c$ is equivalent to Larmor frequency $\omega_L$, because ferrimagnetism in YIG results from electronic spin, so that Eq.~\eqref{eq:cyclotron_freq} is applicable with $\mathcal{M}=1$. 
In the proximity of the resonance, $\omega\approx\omega_0$, one may obtain expressions identical to Eqs.~(6) and~(7) of the main article, with $1/\tau=\alpha\omega_0$ and $\mathcal{A}=\omega_M\omega_0$.

%The expansion leads us to the lengthy expression for the focal length
%\begin{equation}
%\begin{split}
%    f_\pm = &\frac{R_c^2\left[ (1+\alpha^2)^2\omega^4+2(\alpha^2-1)\omega^2\omega_0^2+\omega_0^4 \right]^2}{d\omega_0\omega_M\mathcal{H}_\pm},\\
%    \mathcal{H}_\pm = &\mp2(-1+2\alpha^2+3\alpha^4)\omega^5\omega_0\\ &\mp4(\alpha^2+1)\omega^3\omega_0^3 \pm2\omega\omega_0^5\\ 
%    &+\varepsilon [ (\alpha^2-1)(\alpha^2+1)^2\omega^6 \\
%    &+(1+10\alpha^2+\alpha^4)\omega^4\omega_0^2\\ &-(\alpha^2-1)\omega^2\omega_0^4-\omega_0^6 ].\\
%\end{split}{}
%\end{equation}{}

\subsubsection{Graphene}

\begin{figure}[t]
\centering
\includegraphics[width=\linewidth]{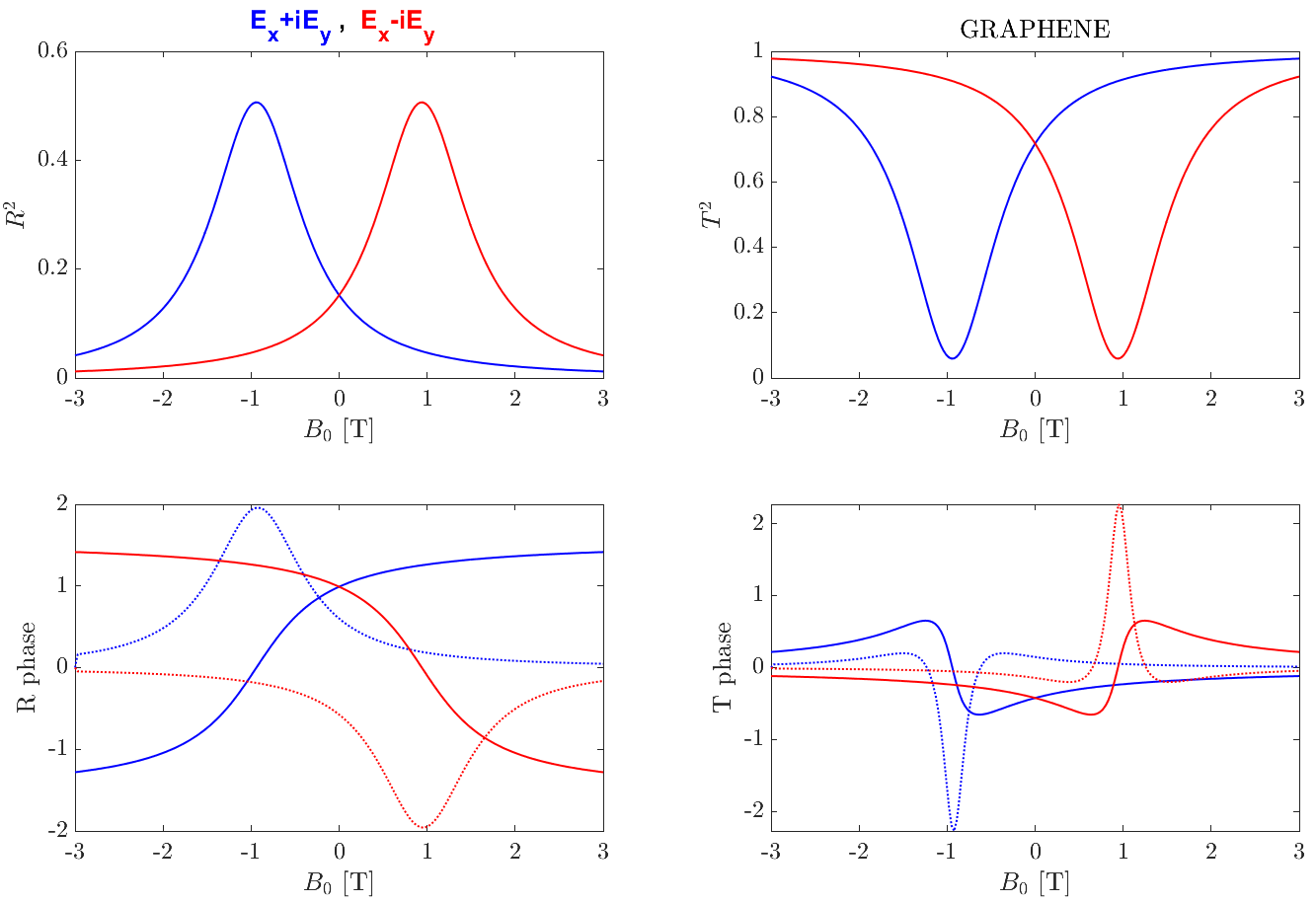}
\caption{(colour online) Cyclotron resonance transition in graphene w.r.t. different circular polarisation components. In bottom panels, dotted curves are the derivatives of phases. Peak positive value of each is a working point for a lens.}
\label{fig:sup_graphene}
\end{figure}

Drude-like model for magnetised graphene is written in terms of conductivity as follows~\cite{Tymchenko2013,Ferreira2011}
\begin{equation}
\begin{split}
    \hat\sigma &= 
    \begin{pmatrix}
        \sigma_{xx} & -i\sigma_{xy} & 0 \\
        i\sigma_{xy}    & \sigma_{yy}   & 0 \\
        0   & 0 & \sigma_{zz} \\
    \end{pmatrix},\\
    \sigma_{xx} = \sigma_{yy} &= \frac{q^2 |\mu_c|}{\pi \hbar^2} \frac{i(\omega-i/\tau)}{(\omega-i/\tau)^2-\omega_c^2}, \\
    \sigma_{xy} &= \frac{q^2 |\mu_c|}{\pi \hbar^2} \frac{\omega_c}{(\omega-i/\tau)^2-\omega_c^2}. \\
\end{split}
\end{equation}
Here, $\mu_c$ is the chemical potential and $\hbar$ is the reduced Planck constant. Isotropic component $\sigma_{zz}$ is of no further interest.
Limiting to intraband transitions only, we follow the transformation $\hat\varepsilon_{\mathrm{eff}} = (4\pi i/\omega d)~\hat\sigma$ and find the elements of the effective permittivity tensor to read [compare to Eq.~(8) in the article]
\begin{equation}\label{eq:eps_graphene}
\begin{split}
    1+\varepsilon_\perp &= \frac{-2 \alpha_0 |\mu_c|}{\pi \hbar}~ \frac{\lambda}{d}~\frac{(\omega-i/\tau)}{(\omega-i/\tau)^2 - \omega_c^2}, \\
    \beta &= \frac{2 \alpha_0 |\mu_c|}{\pi \hbar}~ \frac{\lambda}{d}~\frac{\omega_c}{(\omega-i/\tau)^2 - \omega_c^2}.    \\
\end{split}
\end{equation}
Here $\alpha_0$ is the fine structure constant and $\lambda=2\pi c/\omega$ is the free-space wavelength, $c$ is the speed of light. $\omega_c$ is given by Eq.~\eqref{eq:cyclotron_freq} with a variable $\mathcal{M}=m_eV_F^2/|\mu_c|$, $V_F$ being the Fermi velocity. %, $\mathcal{M}\approx5.69/|\mu_c|$ for $\mu_c$ taken in eV.
Upon series expansion, the focal length takes on the form of Eq.~(9) of the article. Corresponding constant is combined with thickness $d$ and removes it from the expression for $f_\pm$, $\mathcal{A}d=2\alpha_0 |\mu_c|/ \hbar=1.95\cdot10^{19}$~[eV$^{-1}$s$^{-2}$]$\cdot|\mu_c|$~[eV].
An important feature of graphene-based OML is that only one polarisation component is focused resonantly, see Fig.~\ref{fig:sup_graphene}. Thus, it allows for selective focusing of one polarisation or determining the polarisation content of incident light.

\subsubsection{Semiconductor}

\begin{figure}[t]
\centering
\includegraphics[width=\linewidth]{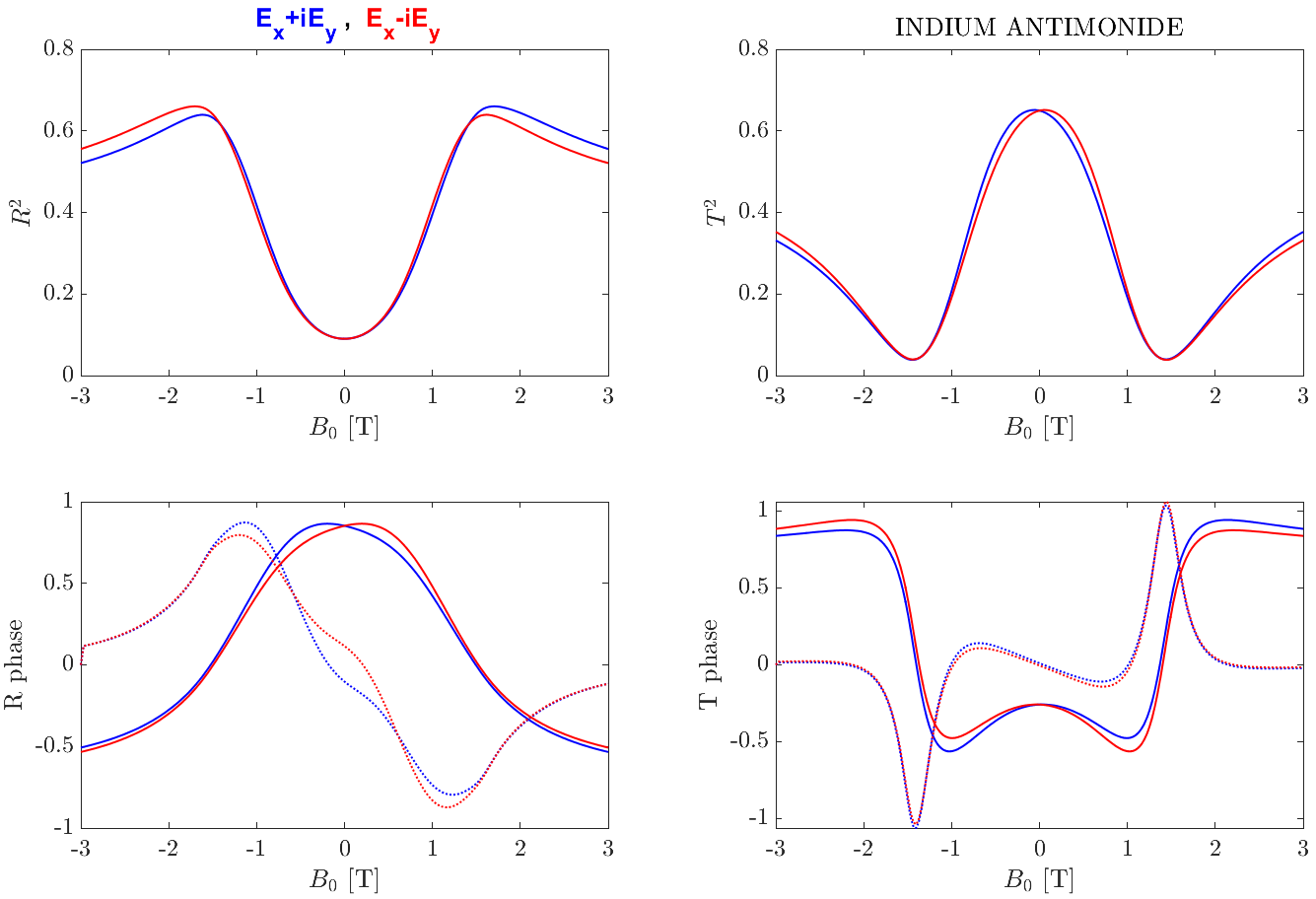}
\caption{(colour online) Cyclotron resonance transition in InSb w.r.t. different circular polarisation components. In bottom panels, dotted curves are the derivatives of phases. Peak positive value of each is a working point for a lens.}
\label{fig:sup_InSb}
\end{figure}

Magnetised semiconductors acquire the tensor form [Eq.~(1) in the article] of dielectric permittivity~\cite{Gibson1995}, with the elements identical to those given by Eq.~(8) of the article 
\begin{equation}\label{eq:eps&beta_semicond}
\begin{split}
    1+\varepsilon_{\perp} &= \varepsilon_\infty - \frac{\varepsilon_\infty \omega_p^2~(\omega - i/\tau)}{\omega \left[ (\omega - i/\tau)^2 - \omega_c^2 \right]},\\
    \beta &= \frac{\varepsilon_\infty \omega_c~\omega_p^2}{\omega \left[ (\omega - i/\tau)^2 - \omega_c^2 \right]}.
\end{split}
\end{equation}
For InSb, $\tau=3.1$~ps, $\varepsilon_\infty=15.68$, $\omega_p=2.43$~THz, and $\mathcal{A}=\varepsilon_\infty \omega_p^2 =9.26\cdot10^{25}$~s$^{-2}$. Upon series expansion, the focal length takes on the form of Eq.~(9) of the article. Unlike in graphene, $\mathcal{M}$ is a constant defined by the process used for manufacturing of the sample. In Table~I (main article), $\mathcal{M}=50$ is assumed. Similarly to ferrites, both polarisations are focused with different effective permittivities, see Fig.~\ref{fig:sup_InSb}.

\subsubsection{Array of magnetic nanoparticles}

From the Eq.~(9) in the article one can see that the focusing effect declines with the frequency increase. Yet it is still present at optical frequencies (near-infrared to visible). A periodic array of ferromagnetic nanoparticles (e.g. disks or pillars made of TbCo~\cite{Ciuciulkaite2020}) allows one to achieve a certain degree of focusing.
Analytical calculations are very limited in this case, while numerical are demanding in computation power. We estimate the efficiency of such an array using a simplified full-wave numerical model. In the model, we sweep over the values of $B_0$ and determine the phase derivative $d\phi/dB_0$. In comparison with InSb (Fig.~\ref{fig:sup_InSb}), it turns out to be a factor of 100 smaller, which is roughly the frequency ratio, $\omega_\mathrm{NIR}=(2\pi)~300$~THz for 1~$\mu$m wavelength.

\subsubsection{Astrophysical plasma}

In outer space there exist directed microwave sources such as cyclotron radiation in the magnetosphere of white dwarfs and pulsars~\cite{zheleznyakov2012radiation,Treumann2006} or maser-like emission in the atmosphere of stars belonging to asymptotic giant branch~\cite{Vlemmings2003}. Extremely high magnetic fields occur nearby pulsars and white dwarfs~\cite{zheleznyakov2012radiation}. These fields are \textit{non-uniform}. Hence, low density plasma nearby stellar objects with high magnetic fields may cause wavefront transformation and affect the perceived position of the source. Formulae given by Eqs.~\eqref{eq:eps&beta_plasma}-\eqref{eq:result_silver_film} are applicable, although with caution.
Astrophysical plasma is often approximated as collisionless, $\tau \rightarrow \infty$. Alternatively, $1/\tau \ll \omega$ and $\omega_0$.
Thus, for quadratic magnetic fields
\begin{equation}\label{eq:result_astro}
    f_\pm \approx \mp \frac{R_c^2}{d\omega_p^2} \frac{\omega}{\omega_0} (\omega \mp \omega_0)^2,
\end{equation}
which may be enough to make the source appear to be at a different distance. %Natural magnetic fields typically have dipole distributions, but the objects themselves are not transparent. However, there are directed sources that may exhibit this effect. For instance, cyclotron radiation in the magnetosphere of white dwarfs and pulsars~\cite{zheleznyakov2012radiation,Treumann2006} or maser-like emission in the atmosphere of stars belonging to asymptotic giant branch~\cite{Vlemmings2003}.
We would like to stress that \textit{any} non-uniformity of the magnetic field over plasma gives a wavefront transformation. In most cases, it would act as aberrations and increase divergence of light.

\subsection{Image formation by Optical Magnetic Lens}

To quantitatively characterise the focusing effect of the OML, we calculate the standard parameters of the focused optical beam: position of a new waist of the  beam and the beam size at the waist. First, we compute how the beam size changes due to OML attenuation.
The inhomogeneous attenuation coefficient $a_\pm = \log\bigl(|\mathcal{T}_\pm|\bigr)$ modifies the size of a new waist $w$ as
\begin{equation}\label{eq:waist_renorm}
    \frac{1}{w^2} = \frac{1}{w_0^2} + \frac{\bigl|a_\pm''\bigr|}{2 R_c^2}, 
\end{equation}
so that the beam size is reduced due to attenuation. 
Hence, the lens equation~\cite{Self1983,Saleh2019} for Gaussian optical beams connecting the position of an object $s$ and image $s'$ (for real image $s'>0$ ) is modified to read 
\begin{equation}\label{eq:new_waist_position}
    \frac{1}{s + z_R^2/(s-f)} + \frac{w/w_0}{s'} = \frac{1}{f}, 
\end{equation}
where $z_R = \pi w_0^2/\lambda$ is the Rayleight length.
Eq.~\eqref{eq:new_waist_position} shows that the image appears closer as compared to the case when the lens attenuation is zero and $w=w_0$.  
The beam size, $w'$, at the new waist position $s'$ is
\begin{equation}\label{eq:new_waist_size}
    w' = wf/\sqrt{(s-f)^2 + z_R^2}
\end{equation}
and depends on the renormalised beam size $w$.

\subsection{Fine tuning of the focal length at the resonant frequency by using two coils}

\begin{figure}[bt]
\centering
\includegraphics[width=\linewidth]{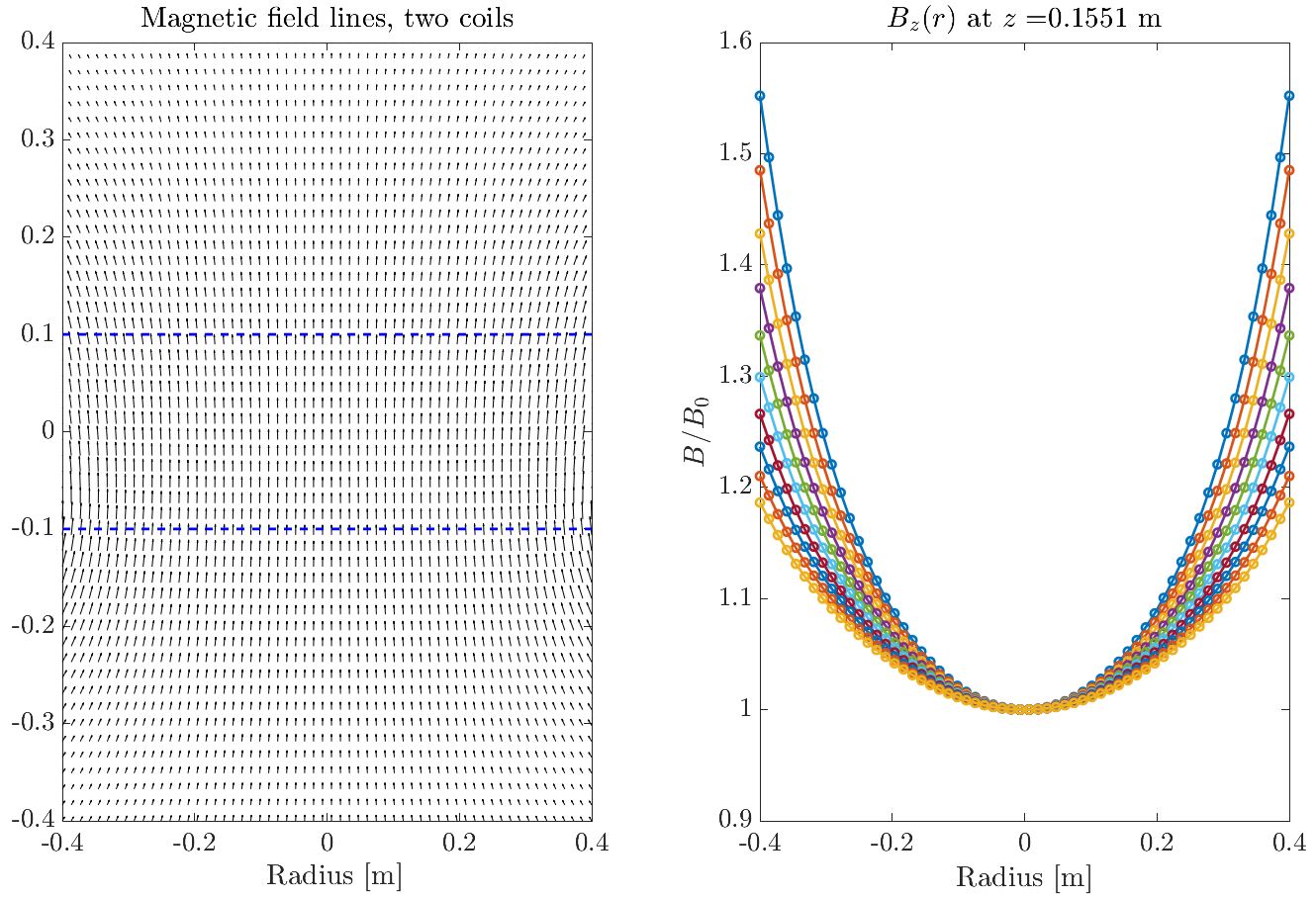}
\caption{(colour online) \textbf{Left panel}: distribution of the magnetic field generated by two coils located at the dashed lines. \textbf{Right panel}: Transverse distributions of the magnetic field $B/B_0=1+r^2/R_c^2$ given by different current ratios $I_1/I_2$ in the coils.}
\label{fig:sup_tuning}
\end{figure}

Optical Magnetic Lens can be tuned precisely by controlling the current ratio of two coils, see Fig.~\ref{fig:sup_tuning}. Importantly, the rate of retuning is limited only by the capabilities of power supplies that feed the coils. The plots in the figure were generated by direct integration of Biot-Savart law. In the center ($y=0$ in the left panel), the field is most uniform, solenoid-like. Thus, an optimal point to locate the OML is slightly out of the coils, where transverse curvature of the magnetic field becomes profound. In the right-hand-side panel one may see that different values of $I_1/I_2$ provide different values of $R_c$.

\subsection{Derivation of the dielectric permittivity tensor in a non-uniform magnetic field}

We derive the tensor of dielectric permittivity of plasma, $\hat\varepsilon$, in a non-uniform magnetic field $\vec B(r)$ from the first principles, namely: (i) \textit{microscopic} Maxwell's equations for the electric and magnetic vectors $\vec E$ and $\vec B$; (ii) the Newton-Lorentz equation of motion of charge carriers in a thin layer; and (iii) the microscopic current in the form of the Klimontovich distribution. For simplicity, we consider the case of electrons in a plasma layer and a monochromatic wave.

Let us start with the motion of charge carriers in combined non-uniform fields (Cartesian coordinates)
\begin{equation}\label{Eq:motion_full}
    \ddot{\vec{R}} = \frac{q}{m} \vec E(\vec R) \mathrm{e}^{i\omega t} + \frac{q}{mc} \left[ \dot{\vec{R}} \times \vec B (\vec R) \right],
\end{equation}
where $\vec{R}(t)$ is the instantaneous position of the charge carrier, $t$ is the time, $\vec E(\vec R)$ is the electric field of an incident light wave of frequency $\omega$, $\vec B(\vec R)$ is the external static non-uniform magnetic field, $c$ is the speed of light, $q$ and $m$ are the charge and mass of the particle respectively. In order to solve it, we expand it into series and thus split the motion into slow and fast components $\vec{R} = \vec{r} + \vec{\xi}$, $\vec{r}$ being a coordinate with a characteristic frequency reaching towards zero, while $\vec\xi$ oscillates with a frequency close to $\omega$. We point out that only the fast component is radiative. The equation of motion for the fast component reads
\begin{equation}
    \ddot{\vec{\xi}} = \frac{q}{m} \vec E(\vec r) \mathrm{e}^{i\omega t} + \frac{q}{mc} \left[ \dot{\vec{\xi}} \times \vec B (\vec r) \right],
\end{equation}
where dependence of $\vec r$ on time can be neglected. With an ansatz $Y = \dot{\vec{\xi}}$, this equation reduces to a non-homogeneous system of differential equations of the first order $\dot Y - \hat A Y = F(t)$, where
\begin{equation}
    \hat A(\vec r) = 
        \begin{pmatrix}
         0  & -\omega_c^z & \omega_c^y \\
         \omega_c^z   &  0    & -\omega_c^x \\
         -\omega_c^y  & \omega_c^x & 0
    \end{pmatrix}, \quad F(\vec r,t) = \frac{q}{m} \vec E(\vec r) \mathrm{e}^{i\omega t}
\end{equation}
and $\vec\omega_c(\vec r) = q\vec B(\vec r)/mc$. Finding a general solution by variation of parameters is straightforward, but tedious, so here we consider only one specific case when $\vec B=(0,0,B_z)$ and $\omega_c=\omega_c^z$. Then, the solution for the fast component of acceleration of charge carriers reads
\begin{equation}\label{Eq:sol_fast_oscillation}
    \dot Y = \ddot{\vec{\xi}} = \frac{q}{m} \mathrm{e}^{i\omega t}
    \begin{bmatrix}
        \frac{\omega^2 E_x(\vec r) + i\omega\omega_c(\vec r) E_y(\vec r)}{\omega^2-\omega_c^2(\vec r)} \\
        \frac{\omega^2 E_y(\vec r) - i\omega\omega_c(\vec r) E_x(\vec r)}{\omega^2-\omega_c^2(\vec r)} \\
        E_z(\vec r)
    \end{bmatrix},
\end{equation}
where one can explicitly see the rise of polarisation mixing. Note that this radiative acceleration of charges contains dependence on the slow macroscopic coordinate $\vec r$.

Let us now turn to the slow component of motion manifested as particle drift. In non-uniform magnetic fields, charged particles experience slow drift along the axis transverse to both the field and the field gradient~\cite{jackson2007classical}. In non-uniform electric fields, such as the field of a Gaussian beam, particles are subject to ponderomotive drift from the region of strong field towards weaker field (away from the beam axis). Thus, the slow part of the equation of motion reads
\begin{equation}\label{Eq:slow_component}
    \ddot{\vec{r}} = \frac{q}{m} (\vec{\xi} \cdot \vec{\nabla}) \vec E(\vec r) \mathrm{e}^{i\omega t} + \frac{q}{mc} \left[ \dot{\vec{r}} \times \vec B (\vec r) + \dot{\vec{\xi}} \times (\vec{\xi} \cdot \vec{\nabla}) \vec B (\vec r) \right].
\end{equation}
Here, $(\vec{\xi} \cdot \vec{\nabla})$ is a scalar differential operator sometimes called the directional derivative, $\vec{\nabla}=(\partial/\partial x,\partial/\partial y,\partial/\partial z)$ and $\vec\xi = \mathrm{e}^{i\omega t} (x_0,y_0,z_0)$ is a value of the fast coordinate that can be found by integrating Eq.~\eqref{Eq:sol_fast_oscillation}. To solve the Eq.~\eqref{Eq:slow_component}, one can time-average the terms that depend on $\vec{\xi}$. Again, a straightforward, but tedious process that we omit here. An example of such procedure applied to the electric field term can be found, for instance, in Ref.~\cite{usikov1988nonlinear}.

Consider the non-uniform magnetic field given by $B_0 p(r)$ and a Gaussian incident beam $|\vec E| \sim \mathrm{e}^{-|r_\perp|^2/2w_0^2}$.
From the time-averaged equation, it is possible to find the drift velocity. 
The interplay of electric and magnetic drift terms melts down to comparing the characteristic sizes of their profiles; namely, the beam waist size $w_0$ and the magnetic field curvature $R_c$.
Both when $w_0 \gg R_c$ and $w_0 \approx R_c$, drift velocities have similar magnitude and opposite signs, which results in negligible net drift.
% Along the radius on the plasma layer surface, the electric field-driven term decays fast, following the Gaussian profile. If the beam waist is small enough,
If $w_0 \ll R_c$, the magnetic field-driven term dominates over the ponderomotive drift. However, the drift direction given by the axisymmetric magnetic field is tangential to the transverse coordinate $r_\perp$. Thus, non-uniformity of the field drives particles into slow spirals around the $z$-axis without critical effects on the concentration.

To check the consistency of this result, we solve the equations of motion [Eq.~\eqref{Eq:motion_full}] numerically. The obtained numerical solution confirms the analytical result. In the dimensionless form, the equations depend on a ratio $E_0/B_0$. Greatly increasing this ratio does not change the qualitative behavior of the system, but increases the area occupied by it (a possible limit by the size of the sample).

Having established the absence of charge density disturbance, we finally assume a hydrodynamic current in the form of Klimontovich distribution
\begin{equation}\label{Eq:Klim_current}
    \vec j = qn \vec v = qn \dot{\vec{\xi}}.
\end{equation}
From microscopic Maxwell's equations, electromagnetic wave equation follows, with the source term given by the current in Eq.~\eqref{Eq:Klim_current}
\begin{equation}
    \left( \Delta - \frac{1}{c^2} \frac{\partial^2}{\partial t^2} \right) \vec E(\vec r,t) = \frac{4 \pi}{c^2} \dot{\vec{j}} = \frac{4 \pi qn}{c^2} \ddot{\vec{\xi}}.
\end{equation}
Using the Eq.~\eqref{Eq:sol_fast_oscillation}, we obtain
\begin{equation}
    \left( \Delta + \frac{\omega^2}{c^2} \right) 
    \begin{bmatrix}
        E_x(\vec r) \\ E_y(\vec r) \\ E_z(\vec r)
    \end{bmatrix}
        = \frac{\omega_p^2}{c^2} 
    \begin{bmatrix}
        \frac{\omega^2 E_x(\vec r) + i\omega\omega_c(\vec r) E_y(\vec r)}{\omega^2-\omega_c^2(\vec r)} \\
        \frac{\omega^2 E_y(\vec r) - i\omega\omega_c(\vec r) E_x(\vec r)}{\omega^2-\omega_c^2(\vec r)} \\
        E_z(\vec r)
    \end{bmatrix},
\end{equation}
where $\omega_p = \sqrt{4\pi q^2 n/m}$ is the plasma frequency. The right-hand-side of this equation (source term) can be easily included on the left as a dielectric permittivity $\varepsilon$. The presence of imaginary cross-terms there indicates that it has a tensor form. Upon equating corresponding matrix products, one can find the permittivity tensor to read
\begin{equation}
    \begin{split}
    &\hat\varepsilon =
    \begin{pmatrix}
        1+\varepsilon_\perp & -i\beta & 0 \\
        i\beta & 1+\varepsilon_\perp & 0 \\
        0 & 0 & \varepsilon_\parallel
    \end{pmatrix}, \quad
    \varepsilon_\perp = \frac{-\omega_p^2 (\omega-i/\tau)}{\omega \left[ (\omega-i/\tau)^2 -\omega_c^2(\vec r)\right]} \\
    &\varepsilon_\parallel = 1-\frac{\omega_p^2} {\omega(\omega-i/\tau)}, \quad \beta = \frac{\omega_p^2 \omega_c(\vec r)}{\omega \left[ (\omega-i/\tau)^2 -\omega_c^2(\vec r)\right]}.
    \end{split}
\end{equation}
Here, we included the phenomenological absorption represented by the relaxation time $\tau$.
Thus, we have shown that under non-uniform magnetic fields the dielectric permittivity tensor for optical beams retains its form while acquiring a coordinate dependence given by the applied field.

%==========================================

\end{document}